\documentclass[12pt]{article}
\usepackage{amsmath,amssymb}
\usepackage{graphicx}
\usepackage{subfigure}
\usepackage[pdftex]{hyperref}
\DeclareGraphicsExtensions{.eps,.bmp,.wmf,.jpg,.pdf}
\numberwithin{equation}{section}
\def\be{\begin{equation}}
\def\ee{\end{equation}}

\def\bea{\begin{eqnarray}}
\def\eea{\end{eqnarray}}

\title{Holographic reconstruction of the {\it k}-essence and dilaton models}
\author{L.N. Granda\thanks{ngranda@univalle.edu.co} \, and\  A. Oliveros\thanks{alexogar@univalle.edu.co}\\
Department of Physics, Universidad del Valle\\ A.A. 25360, Cali,
Colombia} 
\date{}
\begin{document}
\maketitle

\begin{abstract}
\noindent We propose an holographic $k$-essence and dilaton models of dark energy. The correspondence between the $k$-essence and dilaton energy densities with the holographic density, allows the reconstruction of the potential and the fields for the $k$-essence and dilaton models in flat FRW background. For the proposed infrared cut-off the reconstruction was made for the two cases of the constant $\alpha$: for $\alpha<1$ the model presents phantom crossing and the reconstruction was made in the region before the $\omega=-1$ crossing for the EoS parameter. The cosmological dynamics for $\alpha>1$ was also reconstructed. The reconstruction is consistent with the observational data.
\end{abstract}

\section{Introduction}
\noindent Many astrophysical data, such as observations of large scale structure \cite{tegmark}, searches for type Ia supernovae \cite{riess}, and measurements of the cosmic microwave background anisotropy \cite{spergel}, all indicate that the expansion of the universe is undergoing cosmic acceleration at the present time, due to some
kind of negative-pressure form of matter known as dark energy (\cite{copeland},\cite{turner}). Although the cosmological observations suggest that the dark energy component is about $2/3$ of the total content of the universe, the nature of the dark energy as well as its cosmological origin remain unknown at present. The simplest candidate for dark energy is the cosmological constant \cite{weinberg}, \cite{padmana}, \cite{sahni} conventionally associated
with the energy of the vacuum with constant energy density and pressure, and with equation of state $w = -1$. The present observational data favor an equation of state for the dark energy with parameter very close to that of the cosmological constant. The next simple model proposed for dark energy is the quintessence ((see \cite{ratra}, \cite{copeland1}, \cite{caldwell}, \cite{zlatev})), an ordinary scalar field
minimally coupled to gravity, with particular potentials that lead to late time accelerated expansion. The equation of state for a spatially homogeneous quintessence scalar field satisfies $w > -1$ and therefore can produce accelerated expansion. This field is taken to be extremely light
which is compatible with its homogeneity and avoids the problem with the initial conditions \cite{copeland}. Besides quintessence, a wide variety of scalar-field models have been proposed to explain the nature of the dark energy. These include $k$-essence models based on scalar field with non-standard kinetic term \cite{armendariz},\cite{chiba}; string theory fundamental scalars known as tachyon \cite{padmana1} and dilaton \cite{gasperini}; scalar field with negative kinetic energy, which provides a solution known as phantom dark energy \cite{caldwell1}. Other proposals
on dark energy include interacting dark energy models \cite{amendola} \cite{yin}, brane-world models \cite{dvali}, \cite{shtanov}, modified theories of gravity known as f(R) gravity, in which dark energy emerges
from the modification of geometry \cite{carroll},\cite{capozziello}, \cite{odintsov}, \cite{sergei1}, and dark energy models involving non-standard equations of state \cite{kamen},\cite{odin} (for a review on above mentioned and other approaches to dark energy, see \cite{copeland}). In all these models of scalar fields, nevertheless, the potential is chosen by hand guided by phenomenological considerations, lacking the theoretical origin.
Although it has not been established a complete theory of quantum gravity, we can try to find out the nature of dark energy according to some facts of quantum gravity, known as the holographic principle (\cite{beckenstein, thooft, bousso, cohen, susskind}). This principle emerges as a new paradigm in quantum gravity and was first put forward by t' Hooft \cite{thooft} in the context of black hole physics and later extended by Susskind \cite{susskind} to string theory. According to the holographic principle, the entropy of a system scales not with its volume, but with its surface area. In the cosmological context, the holographic principle will set an upper bound on the entropy of the universe \cite{fischler}. In the work \cite{cohen}, it was suggested that in quantum field theory a short distance cut-off is related to a long distance cut-off (infra-red cut-off $L$) due to the limit set by black hole formation, namely, if is the quantum zero-point energy density caused by a short distance cut-off, the total energy in a region of size $L$ should not exceed the mass of a black hole of the same size, thus $L^3\rho_\Lambda\leq LM_p^2$. Applied to the dark energy issue, if we take the whole universe into account, then the vacuum energy related to this holographic principle is viewed as dark energy, usually called holographic dark energy \cite{cohen} \cite{hsu}, \cite{li}. The largest $L$ allowed is the one saturating this inequality so that we get the holographic dark energy density.
\begin{equation}\label{eq1}
\rho_\Lambda=3c^2M_p^2L^{-2}
\end{equation}
where $c^2$ is a numerical constant and $M_p^{-2}=8\pi G$.\\
In the work \cite{li} it was pointed out that the infra-red cutoff $L$ should be given by the future event horizon of the universe, in order to provide the EoS parameter necessarily for the accelerated expansion.\\
By other hand, the scalar field models can be seen as the effective models of the underlying theory dark energy and in this sense the scalar field models can be used to describe the holographic energy density as effective theories. The holographic tachyon  have been discussed in \cite{setare1}, \cite{jingfei}. The holographic phantom quintessence and Chaplygin gas models have been discussed in \cite{setare2,setare3} respectively, and an holographic k-essence model have been considered in \cite{rozas}. In all this models, the infra-red cut-off given by the future event horizon has been used for the reconstruction of the potentials. However this cut-off enters in conflict with the causality \cite{li}. Other reconstructing techniques in theories with a single or multiple scalar fields have been worked in \cite{sodintsov}.

In this paper we consider the scalar field models of $k$-essence and dilaton separately as effective models of the underlying theory of dark energy, and will use the correspondence with the holographic
scenario proposed in \cite{granda}, to reconstruct the field and potential for the $k$-essence field, which reproduces the given holographic DE density. We also considered  the dilaton field without potential and reconstructed the scalar field. The holographic reconstruction in the same scenario for the quintessence and tachyon models has been presented in \cite{granda2}.

\section{The model}
\noindent Let us start with the proposal for the holographic density as given in \cite{granda}
\begin{equation}\label{eq2}
\rho_{\Lambda}=3M_p^2\left(\alpha H^2+\beta \dot{H}\right)
\end{equation}
where $H=\dot{a}/a$ is the Hubble parameter and $\alpha$ and $\beta$ are constants which must satisfy the restrictions imposed by the current observational data. Besides the fact that the underlying origin of the holographic dark energy is still unknown, the inclusion of the time derivative of the Hubble parameter may be expected as this term appears in the curvature scalar (see \cite{gao}), and has the correct dimension. This kind of density may appear as the simplest case of more general $f(H,\dot{H})$ holographic density in the FRW background. This proposal also avoids the coincidence problem as the expression for the holographic density contains a term which track the matter and radiation epochs. Using Eq. (\ref{eq2}), we write the Friedman equation 
\begin{equation}\label{eq3}
H^2=\frac{1}{3M_p^2}(\rho_m+\rho_r)+\alpha H^2+\beta \dot{H}
\end{equation}
where $M_p=(8\pi G)^{-1/2}$ is the Planck mass and $\rho_m$, $\rho_r$ terms are the contributions of
non-relativistic matter and radiation, respectively. This equation can be rewritten in the form (\cite{granda})
\begin{equation}\label{eq4}
\tilde{H}^2=\Omega_{m0}e^{-3x}+\Omega_{r0}e^{-4x}+\alpha \tilde{H}^2+\frac{\beta}{2}\frac{d\tilde{H}^2}{dx}
\end{equation}
in this equation $x=\ln{a}$, $\tilde{H}=H/H_0$, and the subscript $0$ represents the value of a quantity at present ($z=0$). Solving Eq. (\ref{eq4}), we obtain
\begin{equation}\label{eq5}
\begin{aligned}
\tilde{H}^2=&\Omega_{m0}e^{-3x}+\Omega_{r0}e^{-4x}+\frac{3\beta-2\alpha}{2\alpha-3\beta-2}\Omega_{m0}e^{-3x}\\
&+\frac{2\beta-\alpha}{\alpha-2\beta-1}\Omega_{r0}e^{-4x}+Ce^{-2x(\alpha-1)/\beta}
\end{aligned}
\end{equation}
where $C$ is an integration constant. Using the redshift relation $1+z=a_0/a$ with $a_0=1$, the equation (\ref{eq5}) takes the form
\begin{equation}\label{eq6}
\begin{aligned}
\tilde{H}(z)^2=&\Omega_{m0}(1+z)^3+\Omega_{r0}(1+z)^4+\frac{3\beta-2\alpha}{2\alpha-3\beta-2}\Omega_{m0}(1+z)^3\\
&+\frac{2\beta-\alpha}{\alpha-2\beta-1}\Omega_{r0}(1+z)^4+C(1+z)^{2(\alpha-1)/\beta}
\end{aligned}
\end{equation}
the last three terms in \ref{eq6} give the scaled dark
energy density $\tilde{\rho}_{\Lambda}=\frac{\rho_{\Lambda}}{3M_p^2H_0^2}$
\begin{equation}\label{eq7}
\tilde{\rho}_{\Lambda}=\frac{3\beta-2\alpha}{2\alpha-3\beta-2}\Omega_{m0}(1+z)^3\\
+\frac{2\beta-\alpha}{\alpha-2\beta-1}\Omega_{r0}(1+z)^4+C(1+z)^{2(\alpha-1)/\beta}
\end{equation}
and the corresponding pressure density $\tilde{p}_{\Lambda}$ is obtained from the conservation equation $\tilde{p}_{\Lambda}=-\tilde{\rho}_{\Lambda}-1/3 d\tilde{\rho}_{\Lambda}/dx$, and in terms of the redshift is given by
\begin{equation}\label{eq8}
\tilde{p}_{\Lambda}=\frac{2\alpha-3\beta-2}{3\beta}\,C(1+z)^{2(\alpha-1)/\beta}+\frac{2\beta-\alpha}{3(\alpha-2\beta-1)}\,
\Omega_{r0}(1+z)^4
\end{equation}
Considering the equation of state for the present epoch (i.e. at z=0) values of the density
and pressure of the dark energy $\tilde{p}_{\Lambda0}=\omega_0\Omega_{\Lambda0}$ and the Eq. (\ref{eq6}) at the present epoch, we obtain the two equations 
\begin{equation}\label{eq9}
\Omega_{\Lambda}=\frac{3\beta-2\alpha}{2\alpha-3\beta-2}\Omega_{m0}
+\frac{2\beta-\alpha}{\alpha-2\beta-1}\Omega_{r0}+C
\end{equation}
and
\begin{equation}\label{eq10}
\omega_0\Omega_{\Lambda0}=\frac{2\alpha-3\beta-2}{3\beta}\,C+\frac{2\beta-\alpha}{3(\alpha-2\beta-1)}\,
\Omega_{r0}
\end{equation}
from Eqs. \ref{eq9}, \ref{eq10} we can write the constants $\alpha$ and $C$ in terms of  $\beta$, with appropriate values for the parameters $\Omega_{m0}$, $\Omega_{r0}$, $\Omega_{\Lambda 0}$ and $\omega_0$. Once $\tilde{\rho}_{\Lambda}$ and $\tilde{p}_{\Lambda}$ are defined, we can write the expression for the deceleration parameter in terms of the constant $\beta$ (see \cite{granda}). Then we select those values of $\beta$ that give the desired redshift transition according to the astrophysical data . In table I we resume three values for this constants for the case of $\alpha<1$ \cite{granda}.  This values give a negative power-law in the last term in the expression for the holographic density (\ref{eq7}), allowing values of the EoS parameter $w_{\Lambda}$ crossing the phantom barrier and giving rise to a future "Big Rip" singularity \cite{kamionkowski},\cite{nesseris} (for a model with EoS crossing the phantom limit see \cite{xiulian}). As is well known the EoS parameter for the $k$-essence and dilaton models can not cross the phanton (or cosmological constant) limit to $\omega<-1$ values, because of high instability under metric and matter perturbations \cite{vikman}. Therefore the transition from quintessence to phantom phase in this models is not viable, and the holographic reconstruction for $\alpha<1$ will be consistent for $z\geq0$.
\begin{center}\renewcommand{\tabcolsep}{1cm}
\begin{tabular}{|c|c|c|c|}\hline
\multicolumn{4}{|c|}{$\Omega_{m0}=0.27$\ \ \ $\Omega_{\Lambda0}=0.73$\ \ \ $\Omega_{r0}=0$\ \ \ $\omega_0=-1$}\\ \hline\hline
$\beta$ & $z_T$ & $\alpha$ & $C$\\ \hline
0.3 & 0.38& 0.85& 0.55\\ \hline
0.5 & 0.59& 0.93& 0.67\\ \hline
0.6 & 0.69& 0.97& 0.7\\ \hline
\end{tabular}
\end{center}
\begin{center}
\it{Table I} 
\end{center}
By other hand, we can also consider another set of values for $\alpha$, $C$ and $\beta$ as giving in table II. With this data we can reconstruct the $k$-essence and dilaton models in the region $\omega>-1$ for $z\geq-1$. Note that we have taken in this case $w_0=-0.9$ which also gives an adequate red-shift transition and is within the limits set by the different sources of astrophysical data \cite{steen,carolina,tegmark1,turner} (see Figs.1-3 in \cite{granda2}).
\begin{center}\renewcommand{\tabcolsep}{1cm}
\begin{tabular}{|c|c|c|c|}\hline
\multicolumn{4}{|c|}{$\Omega_{m0}=0.27$\ \ \ $\Omega_{\Lambda0}=0.73$\ \ \ $\Omega_{r0}=0$\ \ \ $\omega_0=-0.9$}\\ \hline\hline
$\beta$ & $z_T$ & $\alpha$ & $C$\\ \hline
0.55 & 0.59& 1.01& 0.67\\ \hline
0.65 & 0.68& 1.06& 0.7\\ \hline
0.7 & 0.72& 1.09& 0.72\\ \hline
\end{tabular}
\end{center}
\begin{center}
\it{Table II} 
\end{center}
Note that $\beta$ is the only parameter in this model which needs to be fitted by observational data. Using the values consigned in tables I and II we proceed to the reconstruction of the proposed scalar field models.
\section{Reconstruction of the k-essence model}
In this section, we will discuss the scalar field and potential associated with the $k$-essence model, and will reconstruct them using the correspondence with the holographic principle, in the flat FRW background.
The scalar field model known as $k$-essence is also used to explain the observed late-time acceleration of the universe. It is well known that $k$-essence scenarios have attractor-like dynamics, and therefore avoid the fine tuning of the initial conditions for the scalar field (\cite{chiba},\cite{vitaly}).
This kind of models is characterized by non-standard kinetic energy terms, and are described by a general scalar field action which is a function of $\phi$ and $X=-1/2\partial_{\mu}\phi\partial^{\mu}\phi$, and is given by \cite{damour}
\begin{equation}\label{eq11}
S=\int d^4x\sqrt{-g}p(\phi,X)
\end{equation}
where $p(\phi,X)$ corresponds to a pressure density and usually is restricted to the Lagrangian density of the form
$p(\phi,X)=f(\phi)g(X)$. Based on the analysis of the low-energy effective action of string theory (see \cite{damour} for details) the
Lagrangian density can  be transformed into
\begin{equation}\label{eq12}
p(\phi,X)=f(\phi)\left(-X+X^2\right)
\end{equation}
From the energy momentum-tensor for this Lagrangian density it follows the next expression for the energy density
of the field $\phi$ (see \cite{damour})
\begin{equation}\label{eq13}
\rho(\phi,X)=f(\phi)\left(-X+3X^2\right)
\end{equation}
And the equation of state using (\ref{eq12}) and (\ref{eq13}) is given by
\begin{equation}\label{eq14}
\omega_K=\frac{X-1}{3X-1}
\end{equation}
The condition $1/2<X<2/3$ gives an equation of state $-1<\omega_K<-1/3$ giving rice to accelerated expansion, and the equation of state of the cosmological constant corresponds to $X=1/2$.
In order to establish the correspondence with the holographic model, the $k$-essence energy density and EoS parameter $\omega_{K}$  will be matched to the corresponding holographic energy density and EoS
parameter  $\omega_{\Lambda}=\tilde{p}_{\Lambda}/\tilde{\rho}_{\Lambda}$ (with $\tilde{\rho}_{\Lambda}$ and $\tilde{p}_{\Lambda}$ given by Eqs. (\ref{eq7}),(\ref{eq8}) respectively). This is in complete agreement with the Friedmann equations for the holographic (plus dark matter) and $k$-essence (plus dark matter) models separately. Therefore, from Eq. (\ref{eq14}) the kinetic term $X$ can be written in terms of the holographic quantities. 
\be\label{eq15}
X=\frac{1}{2}\dot{\phi}^2=\frac{\omega_{\Lambda}-1}{3\omega_{\Lambda}-1}
\ee
and replacing $\omega_{\Lambda}$ through $\tilde{\rho}_{\Lambda}$ and $\tilde{p}_{\Lambda}$ (hereafter the contribution of the radiation will be dropped)
\be\label{eq16}
X=\frac{1}{3}\frac{2C\left(2\alpha^2+9\beta^2-4\alpha+9\beta-9\alpha\beta+2\right)-3\beta(3\beta-2\alpha)\Omega_m (1+z)^{3-2(\alpha-1)/\beta}}{2C\left(2\alpha^2+6\beta^2-4\alpha+7\beta-7\alpha\beta+2\right)-\beta(3\beta-2\alpha)\Omega_m(1+z)^{3-2(\alpha-1)/\beta}}
\ee
The behavior of the kinetic term $X$ with respect to $z$ is shown in Fig. 1.
\begin{center}
\includegraphics [scale=0.7]{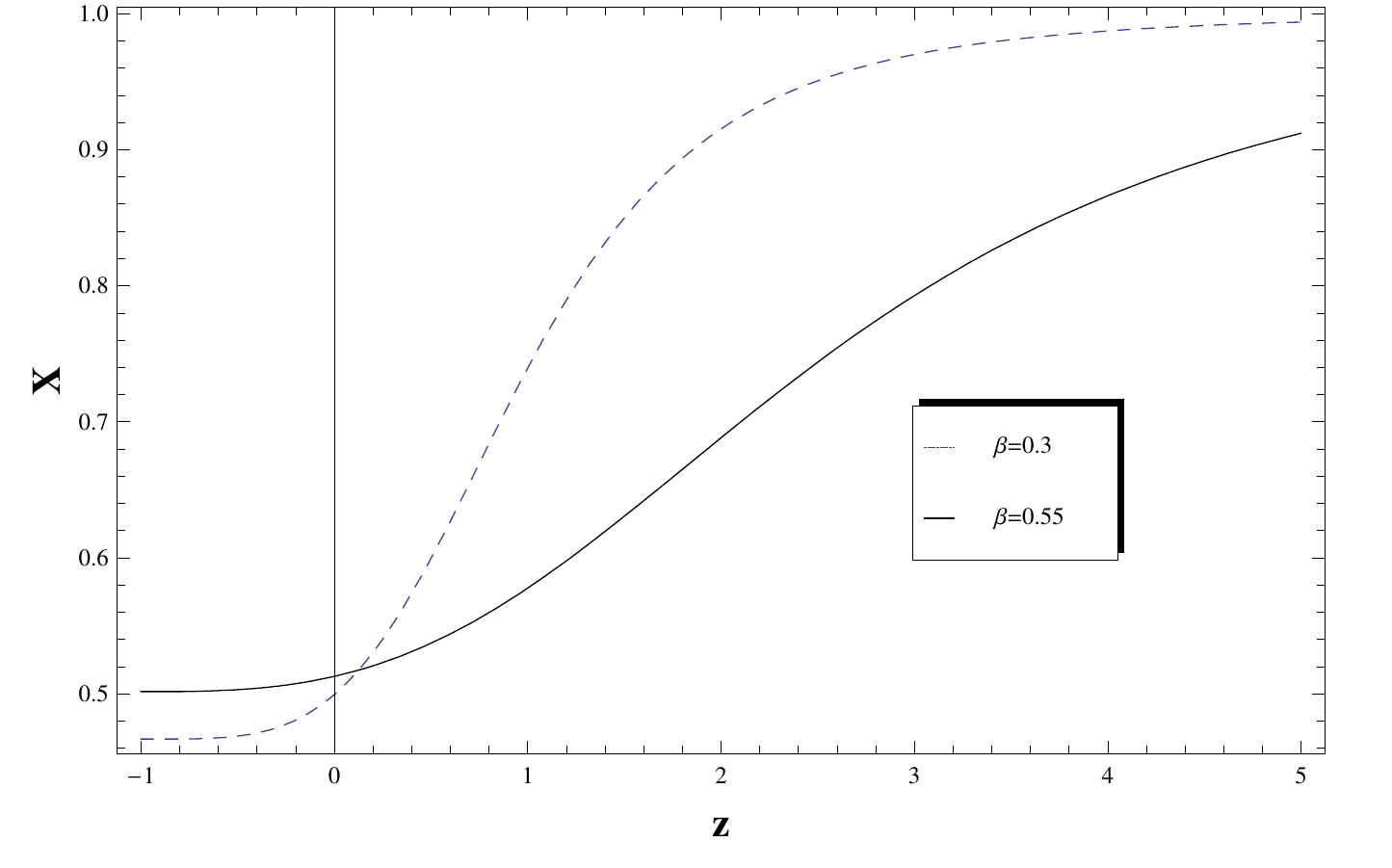}
\end{center}
\begin{center}
Figure 1: The evolution of the kinetic term $X$ in terms of the redshift, for two representative values taken form tables I and II. At high redshift $X\rightarrow1$ and the curves behave as presureless dark matter, and at low redshift $z\rightarrow0$ the $\alpha<1$ ($\beta=0.3$)curve tends to a cosmological constant behavior, and the $\alpha>1$ ($\beta=0.55$) curve tends to slightly higher value. 
\end{center}
Note that in Fig. 1 the $\alpha<1$ curve cross the phantom limit $X=1/2$ which is forbidden for the $k$-essence model. Therefore in this case, the correspondence is consistent for $z\geq0$. The curve for $\alpha>1$ can be plotted for values of $z\geq-1$, as in this case we have no phantom crossing, and at $z\rightarrow-1$ in the future, the cosmological constant  behavior is reached.\\
Taking into account that $H=H_0\tilde{H}$ and turning the time derivative to the redshift variable $z=1/a-1$, one can find from Eq. (\ref{eq15}) the following equation for the field $\phi$ 
\begin{equation}\label{eq17}
\frac{d\phi}{dz}=\mp\frac{1}{(1+z)H_0}\left[\frac{2}{\tilde{H}^{2}}\frac{\tilde{p}_{\Lambda}-\tilde{\rho}_{\Lambda}}{3\tilde{p}_{\Lambda}-\tilde{\rho}_{\Lambda}}\right]^{1/2}
\end{equation}
with $\tilde{H}^2$, $\tilde{\rho_{\Lambda}}$ and $\tilde{p_{\Lambda}}$  given by Eqs. (\ref{eq6}), (\ref{eq7}) and (\ref{eq8}) respectively. 
This equation can not be integrated exactly, but can be plotted numerically for a given interval of $z$. The plot 
of $\phi$ as function of $z$ is shown in Fig. 2. In fact we have plotted $(\phi(z)-\phi(0))$ but this does not affect the shape of the potential.
\begin{center}
\includegraphics [scale=0.7]{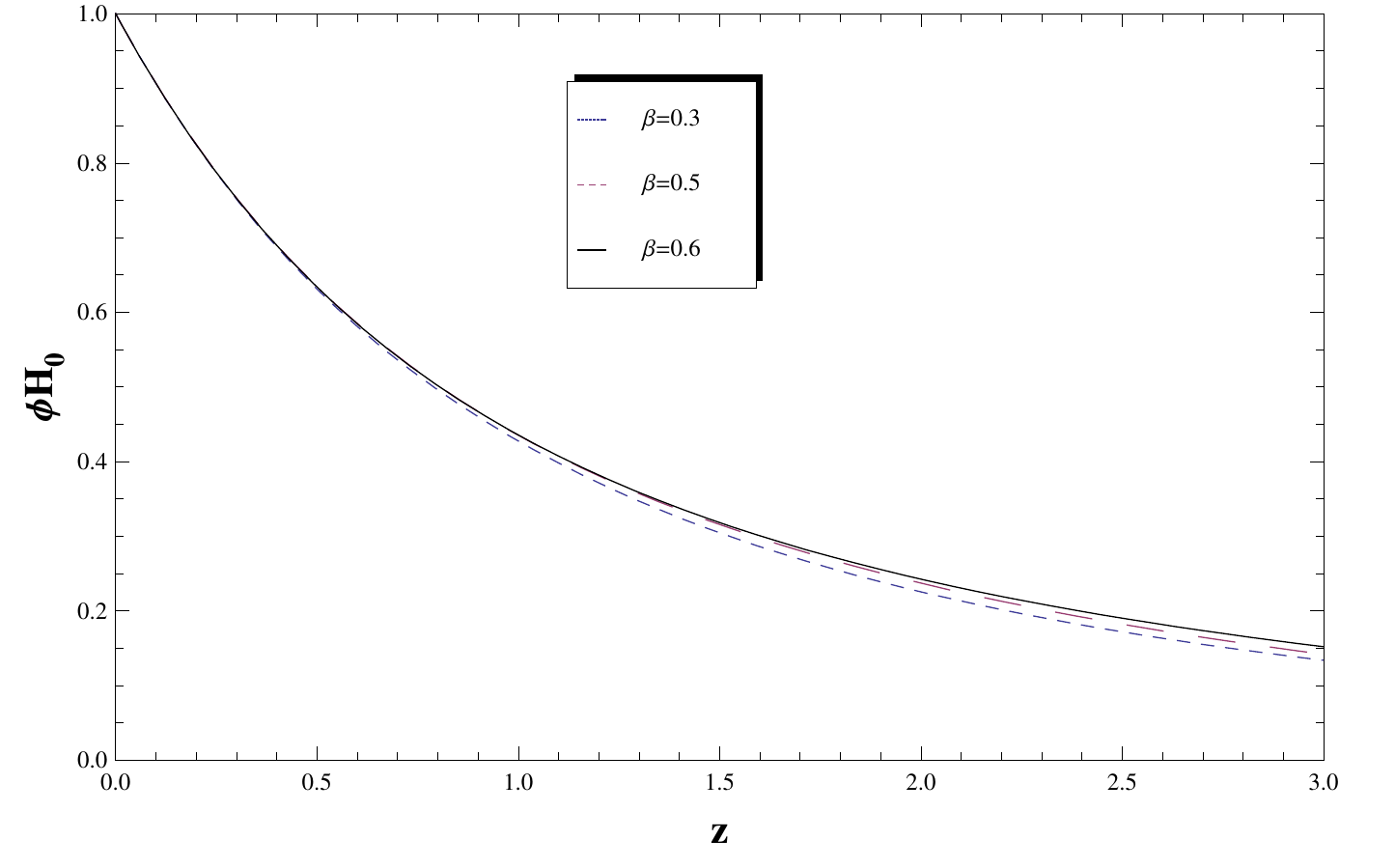}
\end{center}
\begin{center}
Figure 2: The $k$-essence scalar field in terms of the redshift, with the values given in table I. Very similar results are obtained with the values of table II. The $(-)$ sign in Eq. (\ref{eq17}) have been chosen and a displacement in the $H_0\phi$ axis by $1$ have been done.
\end{center}
Note that the field decreases with the increment of $z$, and becomes finite at low redshift. 
From Eqs. (\ref{eq13}) and (\ref{eq7}) it can be obtained an expression for the $k$-essence potential $f(\phi)$ in terms of the redshift $z$
\be\label{eq18}
f(\phi)=3H_0^{2}M_p^{2}\frac{\tilde{\rho}_{\Lambda}\left(1-3\omega_{\Lambda}\right)^{2}}{2\left(1-\omega_{\Lambda}\right)}
\ee
where the Eq. (\ref{eq15}) was used. 
Although is not possible to obtain an analytical expression for the potential in terms of the field, we can numerically plot the behavior of the potential given by Eq. (\ref{eq18}), in terms of the $k$-essence field given by the solution to the Eq. (\ref{eq17}), as is shown in Fig. 3. 

\begin{center}
\includegraphics [scale=0.7]{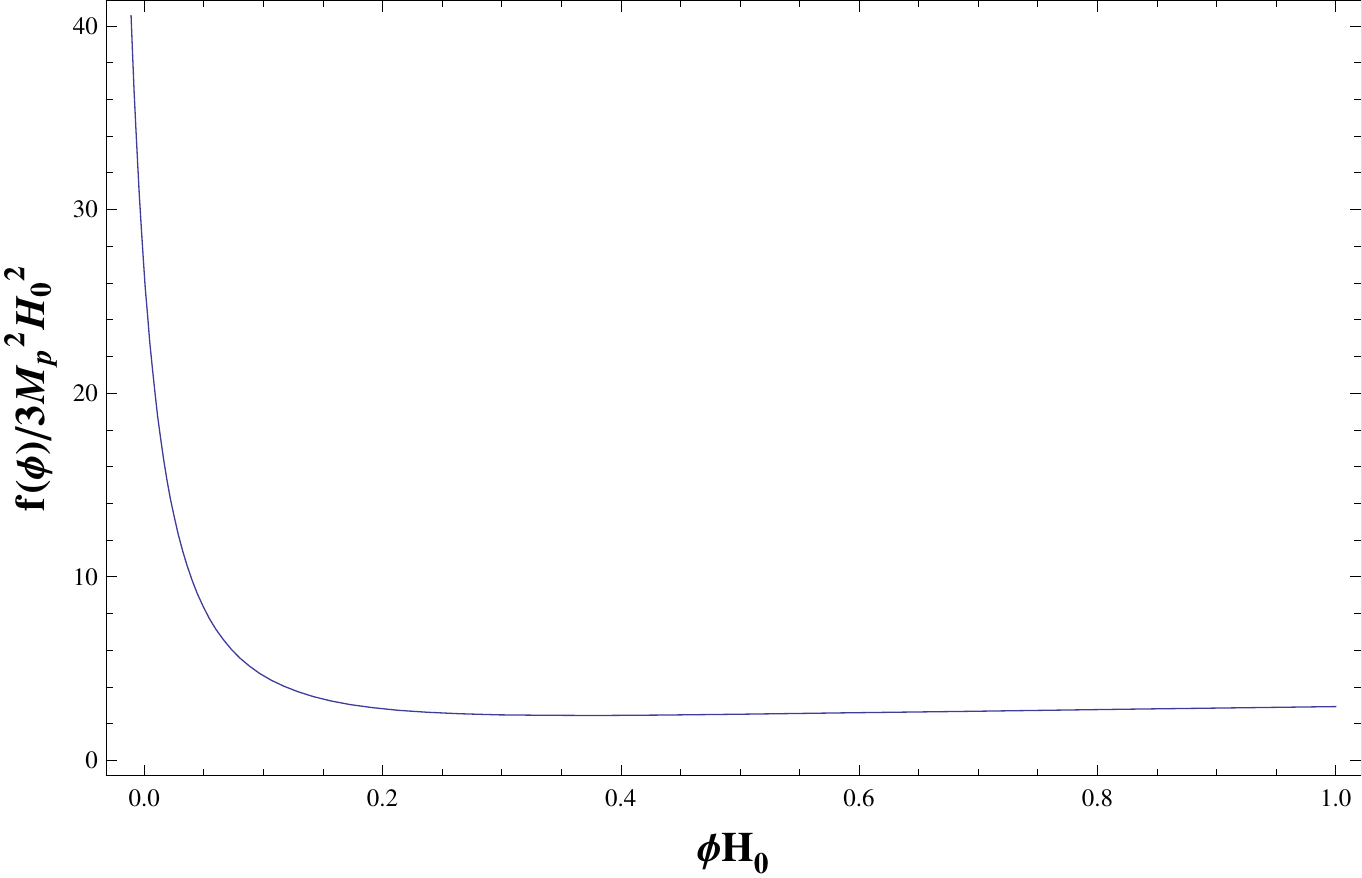}
\end{center}
\begin{center}
Figure 3: The $k$-essence scalar potential in terms of the scalar field, for $\alpha=0.93$ and $\beta=0.5$. Very similar results are obtained with the rest of the values consigned in tables I and II.
\end{center}
Note that the potential at low redshift shows a shape similar to an inverse power-law $f(\phi)\sim 1/\phi^{q}$ with respect to the scalar field. Tough this behavior does not affect the EoS parameter, this kind of potential was considered in the literature and has scaling solution \cite{tsujikawa}.\\

\section{Reconstruction of the dilaton}
In this section we will discuss the reconstruction of the dilaton field, in a flat FRW background. The dilaton field is described by the effective Lagrangian density $p_D(X,\phi)$
\begin{equation}\label{eq20}
p_D(X,\phi)=-X+ce^{\lambda\phi}X^{2}
\end{equation}
where $c$ is a positive constant and $X=-1/2\partial_{\mu}{\phi}\partial^{\mu}\phi$. This model appears from a four-dimensional effective low-energy string action \cite{piazza}
and includes higher-order kinetic corrections to the tree-level action in low energy effective string theory. Writing the Einstein equations with the dilaton field as the source of the energy-momentum tensor, is easy to see that the Lagrangian density (Eq. (\ref{eq20})) actually corresponds to the pressure, $p_D$ of the scalar field, while the energy density is given by (see \cite{piazza})
\be\label{eq21}
\rho_D= 2X\frac{\partial p_D}{\partial X}-p_D=-X+3ce^{\lambda\phi}X^{2}
\ee
with the equation of state parameter
\be\label{eq22}
\omega_D=\frac{cXe^{\lambda\phi}-1}{3cXe^{\lambda\phi}-1}
\ee
In this case the scaling solution corresponds to $Xe^{\lambda\phi}=const.$, which has the solution $\phi\sim\log(t)$. The condition for an accelerated expansion 
gives $1/2<cXe^{\lambda\phi}<2/3$ and the cosmological constant limit corresponds to $cXe^{\lambda\phi}=1/2$.

In order to consider the dilaton field as the effective description of the holographic density, the correspondence between the dilaton energy density and the holographic energy density must be used. This translates into $\rho_D=\rho_{\Lambda}$, giving the equation 
\begin{equation}\label{eq23}
-X+3ce^{\lambda\phi}X^{2}=3M_p^2H_0^2\tilde{\rho}_{\Lambda}
\end{equation}
with $\tilde{\rho}_{\Lambda}$ given by (\ref{eq7}).
The correspondence with the holographic dark energy equation of state ($\omega_D=\omega_{\Lambda}$) gives
\begin{equation}\label{eq24}
cXe^{\lambda\phi}=\frac{\omega_{\Lambda}-1}{3\omega_{\Lambda}-1}
\end{equation}
Using this equation in Eq. (\ref{eq23}) one gets the following equation
\be\label{eq25}
X=\frac{3}{2}H_0^2M_p^2\left(1-3\omega_{\Lambda}\right)\tilde{\rho}_{\Lambda}
\ee
Turning to the redshift variable one can write the equation for the dilaton field as follows
\be\label{eq26}
\begin{aligned}
\frac{d\phi}{dz}=&\mp\frac{\sqrt{3}M_p}{1+z}\left[\frac{\tilde{\rho}_{\Lambda}-3\tilde{p}_{\Lambda}}{\tilde{H}^2}\right]^{1/2}\\
&=\mp\frac{\sqrt{3}M_p}{1+z}\Big[\frac{2C(2\beta-\alpha+1)(3\beta-2\alpha+2)+\beta(2\alpha-3\beta)\Omega_{m0}(1+z)^{3-2(\alpha-1)/\beta}}{C\beta(3\beta-2\alpha+2)+2\beta\Omega_{m0}(1+z)^{3-2(\alpha-1)/\beta}}\Big]^{1/2}
\end{aligned}
\ee
where the definition $\omega_{\Lambda}=\tilde{p}_{\Lambda}/\tilde{\rho}_{\Lambda}$ has been used. 
The integration can be performed exactly, but the analytical expression for $\phi$ is too large. Fig.4 shows the behavior of the dilaton scalar field as function of the redshift. For $\alpha>1$ the behavior is very similar.
\begin{center}
\includegraphics [scale=0.7]{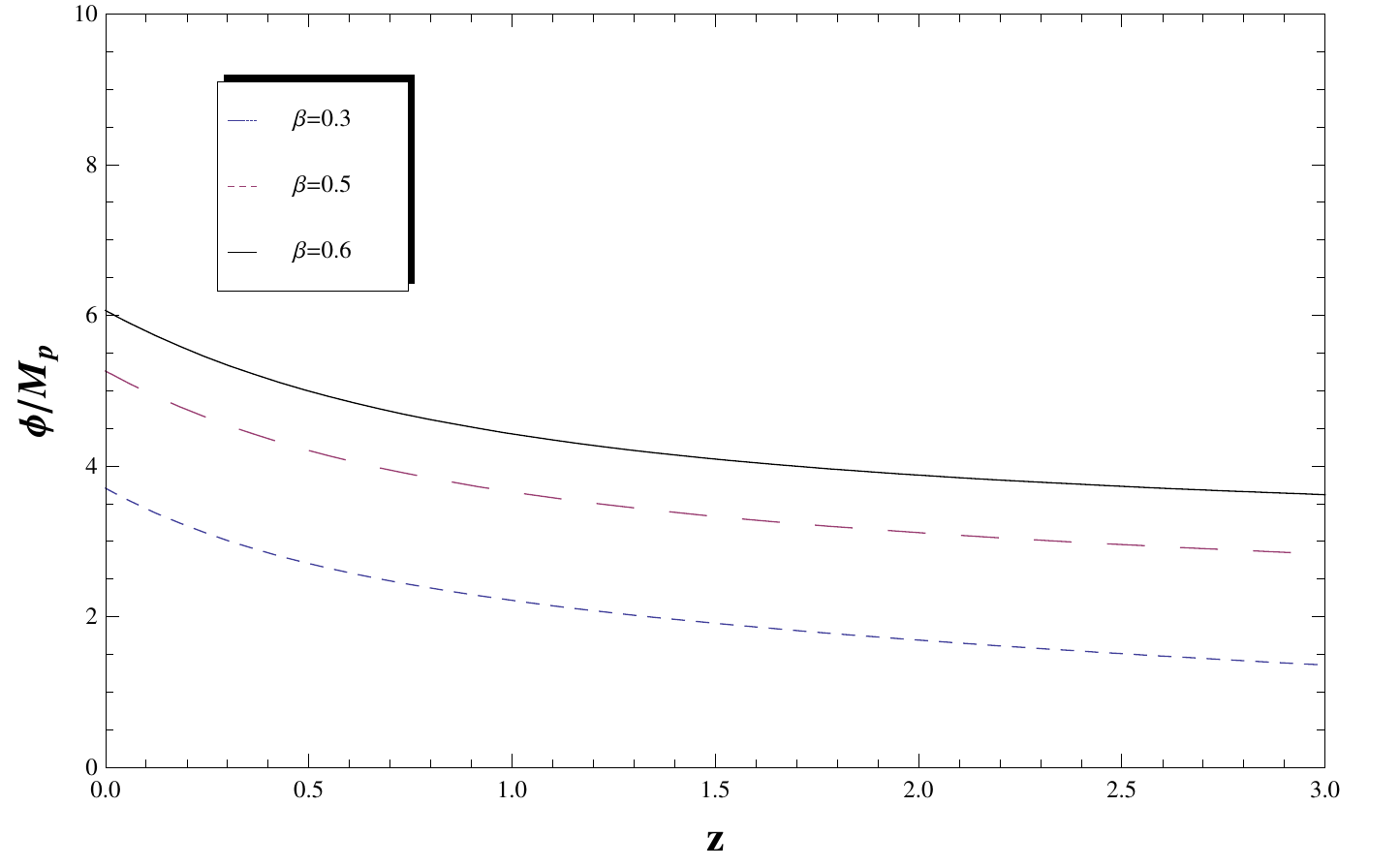}
\end{center}
\begin{center}
Figure 4: The dilaton scalar field as function of the redshift for the values of table I, with the minus sign in Eq. (\ref{eq26}). 
\end{center}

\section{Discussion}
Cosmological scenarios involving a scalar fields appearing in the low-energy effective string theory known
as $k$-essence and dilaton, are intended to explain the late-time acceleration of the universe. In this paper we have studied the $k$-essence and dilaton fields as effective theories that describe the holographic dark energy. In the case of the $k$-essence, the scalar field and the potential have been reconstructed according to the data given in tables I and II. From fig. 1 we see that though the behavior of the the kinetic terms are similar in the region of low redshift, the $\alpha<1$ curve in fig. 1 crosses the phantom barrier which is unphysical for the $k$-essence model, and the $\alpha>1$ curve is more compatible with the model as the limiting case of the phantom crossing does not appear, and tends to a cosmological constant behavior at $z\rightarrow-1$ in the future. Nevertheless, it should be noted that the reconstructed k-esence model for the case $\alpha<1$ (see fig.1), has a continuous transition from the quintessence to phantom phase, despite the problems with instabilities (one possible way out of this problem is to consider the phantom phase as a transient phenomena). 

The correspondence with the holographic density (\ref{eq7}) guarantees that the energy density of $k$-essence and dilaton are subdominant during the matter and radiation epochs, respecting the bounds imposed by the big bang nucleosintesis. One important fact of this reconstruction for both models, is that the equation of state $\omega_{K,D}$ is decreasing at the present epoch towards $\omega = -1$.\\
In summary, we have carried out a reconstruction and analyzed the cosmological evolution of the $k$-essence and dilaton models of dark energy, in the frame of the holographic principle. We used the infrared cut-off for the holographic density, as proposed in \cite{granda}. It can be noted that the reconstruction has been successful in reproducing the main characteristics of this string theory inspired models, relevant for its cosmological dynamics.

\section*{Acknowledgments}
This work was supported by the Universidad del Valle.

\end{document}